\documentclass[amsmath,superscriptaddress,reprint,aps]{revtex4-2}
\usepackage{graphicx}% Include figure files
\usepackage{dcolumn}% Align table columns on decimal point
\usepackage{bm}% bold math
\usepackage{hyperref}% add hypertext capabilities
\usepackage{siunitx}
\usepackage{color, soul}

\setstcolor{red}

\begin{document}
    
	%\preprint{APS/123-QED}
	
	\title{Ultralong pump–probe movies of magnon and phonon dynamics from ultrafast generation to microsecond relaxation}% Force line breaks with \\

     \author{Riku~Shibata}
    \affiliation{Department of Physics, Faculty of Science and Technology, Keio University, Yokohama, 223-8522, Japan}

    \author{Shun~Fujii}
    \affiliation{Department of Physics, Faculty of Science and Technology, Keio University, Yokohama, 223-8522, Japan}

	\author{Shinichi~Watanabe}
	\email[Corresponding author. ]{watanabe@phys.keio.ac.jp}
 	\affiliation{Department of Physics, Faculty of Science and Technology, Keio University, Yokohama, 223-8522, Japan}
	
%	\date{\today}% It is always \today, today,
	% but any date may be explicitly specified
	
\begin{abstract}
The long lifetimes of magnons and phonons make them attractive for information-processing devices, highlighting the importance of visualizing their spatiotemporal dynamics from generation through relaxation. Ultrafast pump–probe spectroscopy is a powerful tool for investigating their early-stage dynamics after impulsive excitation; however, their long-lived nature makes it challenging to comprehensively track their evolution across all relevant time scales while maintaining sufficient temporal resolution. Here, we demonstrate spatiotemporal tracking of magnon and phonon dynamics over more than seven orders of magnitude in time, from 500 femtoseconds to 20 microseconds, using $4 \times 10^7$ sampled time points enabled by the highly precise time base of optical frequency combs. The resulting spatiotemporal movie, consisting of $4.5 \times 10^{5}$ frames, captures their generation, coherent motion, propagation, and relaxation, providing a powerful platform for exploring their full dynamical evolution.
\end{abstract}

	\maketitle

	%\tableofcontents
\section*{Introduction}
\noindent

Magnons and phonons—collective excitations of spin and lattice degrees of freedom, respectively—are key quasiparticles in  next-generation information devices. Magnons can carry magnetic information without charge-current flow and its associated Joule heating \cite{Kajiwara:2010}, while confined phonons provide long-lived channels in elastic media \cite{Wang:2022,Yao:2025}. Magnetoelastic coupling further links these two excitations, enabling the long-distance propagation of hybrid magnon–phonon modes in certain systems \cite{Casals:2020,Zhang:2021,Maezawa:2024}. Directly tracking the ultrafast generation, propagation, and eventual decay of these excitations is therefore essential for elucidating their whole spatiotemporal dynamics and guiding the design of future high-speed spintronic and phononic information devices \cite{Hatanaka:2023,He:2026}. Ultrafast pump–probe imaging provides a powerful means of directly visualizing their early-stage dynamics following impulsive optical excitation. This approach has enabled observations of directional magnon propagation \cite{Satoh:2012,Au:2013,Hashimoto:2017}, crystal-axis-dependent phonon propagation \cite{Sugawara:2002}, and the propagation of coupled magnon–phonon excitations \cite{Ogawa:2015,Hashimoto:2018,Hioki:2019,Hioki:2022}. However, a major limitation remains: although magnons and phonons can persist for several microseconds or longer \cite{Serha:2026,Shao:2019}, conventional ultrafast pump–probe imaging has typically been limited to temporal windows of only several tens of nanoseconds \cite{Hioki:2022}. This limitation primarily arises from the restricted travel range of mechanical delay lines in pump–probe setups. Moreover, the large inertia of the moving mirrors slows delay scanning and limits the number of frames available for constructing a detailed spatiotemporal movie. Consequently, conventional ultrafast pump–probe imaging cannot readily capture dense spatiotemporal movies over the characteristic lifetimes of these quasiparticles.

Recent advances in laser synchronization and timing control have extended ultrafast pump–probe measurements beyond the constraints of conventional mechanical-delay-line systems. Following the pioneering work of Elzinga {\it et al.} \cite{Elzinga:1987a,Elzinga:1987b}, asynchronous optical sampling (ASOPS) employs two electronically synchronized femtosecond pulse trains with slightly different repetition rates to rapidly scan the pump–probe delay without a mechanical delay line \cite{Yasui:2005,Bartels:2006,Krauss:2015,Kim:2018,Asahara:2020,Velsink:2023,Nishikawa:2023}. Although ASOPS enables dense spatiotemporal movies of photogenerated magnons and phonons \cite{Abbas:2014,Nishikawa:2025}, its temporal window is limited to one laser repetition period, typically 1–20 ns for repetition rates of 50–1000 MHz. Efforts to reach the millisecond range have employed kilohertz-repetition-rate amplified lasers with either ASOPS-based approaches \cite{Antonucci:2015,Solinas:2017,Helbing:2023} or electronic trigger-control schemes \cite{Bredenbeck:2004,Yu:2005,Carroll:2009,Domke:2012,Domke:2018,Flory:2023}. However, dense sampling over such long windows is impractical and has not been performed: a 1-ms window sampled at 1-ps intervals contains 10$^{9}$ delay points, requiring 10$^6$ s (11.6 days) for a single time trace at a 1-kHz probe sampling rate. Combining a kilohertz-repetition-rate pump with a megahertz-repetition-rate probe could provide both a wide temporal window and dense temporal sampling; however, synchronizing such disparate pulse trains remains challenging. To date, only stochastic sampling without timing stabilization has been demonstrated\cite{Nakagawa:2016}, and this approach has not yet been extended to spatiotemporal imaging. A reliable and uniformly sampled pump–probe delay axis over an extended temporal window is therefore essential for constructing ultralong spatiotemporal movies.

Here, we demonstrate an ultralong pump–probe movie spanning picosecond-to-microsecond timescales. The 50~kHz repetition rate of the regenerative-amplifier (RA) pump corresponds to a 20~$\mu$s interval between successive pump pulses, thereby defining the observation window. Dense probe sampling is enabled by a 61.6~MHz probe pulse train precisely synchronized with the pump pulse train, yielding a minimum delay increment of 500~fs and up to approximately $4\times10^7$ temporal sampling points within the 20~$\mu$s interval. This capability is achieved using a fully stabilized optical-frequency-comb pair, in which one comb serves as the seed source for the RA. The delay of each probe pulse relative to the corresponding pump pulse is determined with an accuracy of 50~fs, and subsequent data sorting reconstructs time-resolved signals on a uniformly spaced delay grid. The high sampling rate allows each single-scan measurement to be completed in less than 1~s, enabling point-by-point spatial scanning for the construction of detailed spatiotemporal movies. We apply this technique to image optically excited magnon and phonon dynamics in Bi-substituted yttrium iron garnet (Bi-YIG). The resulting movies clearly visualize the impulsive generation, coherent motion, propagation, and relaxation of both magnons and phonons, together with phonon reflection at boundaries, as they evolve across picosecond-to-microsecond timescales. Our approach enables detailed, high-throughput visualization of nonequilibrium dynamics over an ultrabroad temporal window, providing a versatile platform for spatiotemporally mapping diverse excitations in solid-state materials from their generation through relaxation.

\begin{figure*} % Do NOT use \begin{figure*}
	\centering
	\includegraphics[width=1\textwidth]{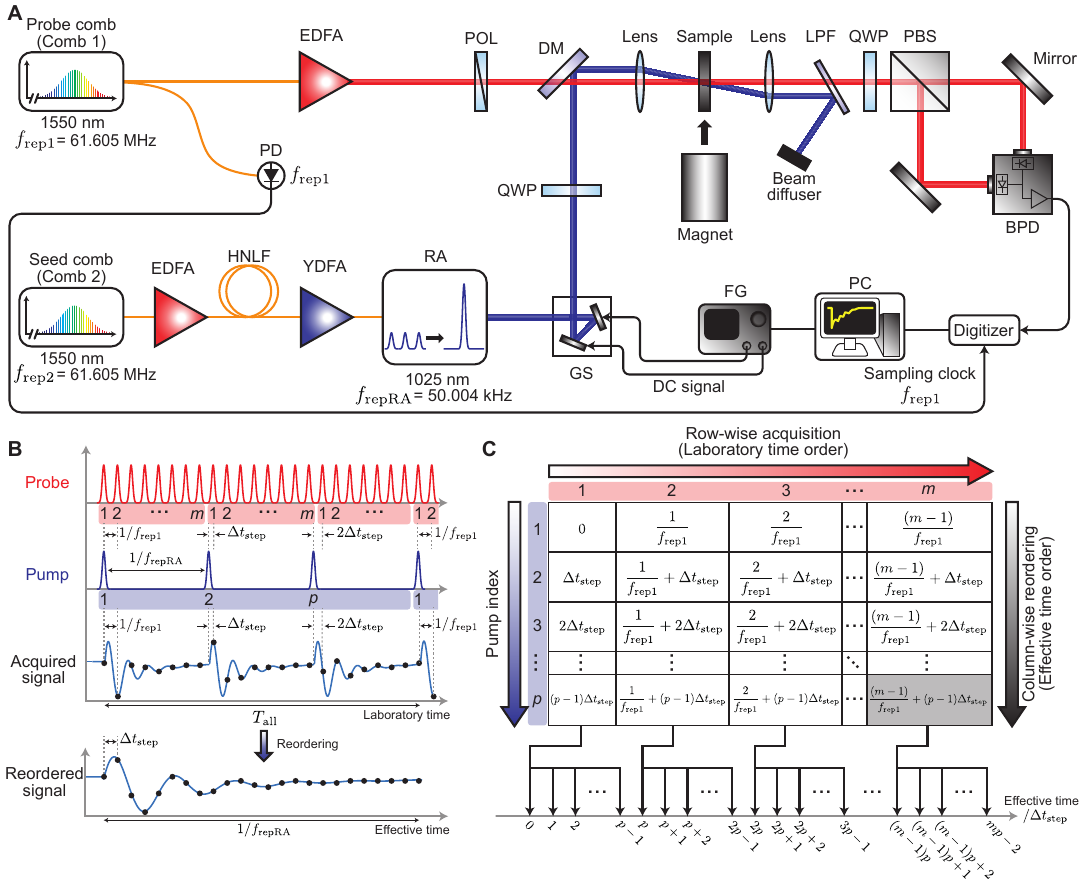} 
\caption{\textbf{Experimental setup and pulse-index reordering principle.}
\textbf{(A)} Schematic of the experimental setup. Comb 1 is amplified and used as the probe comb, and its photodetected repetition-rate signal provides the sampling clock, $f_{\mathrm{rep1}}$, for the digitizer. Comb 2 provides a seed pulse train for the regenerative amplifier, which selects one pulse out of every $m$ seed-comb pulses and generates the amplified pump train at $f_{\mathrm{repRA}}=f_{\mathrm{rep2}}/m$. Pump-induced changes in the probe are measured by balanced detection after polarization analysis. Comb 1 and Comb 2 are fully stabilized with respect to each other, as explained in the text.
\textbf{(B)} Timing concept. The RA pump pulses are separated by $T_{\mathrm{RA}}=1/f_{\mathrm{repRA}}$, and $m$ probe-comb pulses are assigned to one pump period. Owing to the controlled detuning between the two combs, the pump--probe delay shifts by $\Delta t_{\mathrm{step}}$ from one pump cycle to the next. The laboratory-time signal therefore contains samples from different regions of the final delay trace in an interleaved order and is reordered along the pump--probe delay axis, $\tau$.
\textbf{(C)} Matrix representation of the reordering. The data are acquired row-wise in laboratory time, with rows representing successive pump cycles and columns representing the probe-pulse order within one pump period. The row and column numbers shown in the schematic are ordinal labels; using indices $j=0,\ldots,p-1$ and $k=0,\ldots,m-1$, the delay assigned to each matrix element is $\tau_{j,k}=j\Delta t_{\mathrm{step}}+k/f_{\mathrm{rep1}}=(j+kp)\Delta t_{\mathrm{step}}$, where $p$ is the number of pump cycles in one acquisition cycle and $p\Delta t_{\mathrm{step}}=1/f_{\mathrm{rep1}}$. Column-wise deinterleaving, equivalent to sorting the data by the delay index $n=j+kp$, yields the uniformly sampled pump--probe trace $\tau_n=n\Delta t_{\mathrm{step}}$ over one RA period. Note that the bottom-right element in the nominal grid, $\tau_{p-1,m-1}$ (gray-shaded), corresponds to $T_\mathrm{RA}$ and thus coincides with the zero-delay point at the upper-left corner in the subsequent trace.
PD, photodetector; EDFA, erbium-doped fiber amplifier; HNLF, highly nonlinear fiber; YDFA, ytterbium-doped fiber amplifier; RA, regenerative amplifier; POL, polarizer; DM, dichroic mirror; QWP, quarter-wave plate; LPF, long-pass filter; PBS, polarizing beam splitter; BPD, balanced photodetector; GS, galvanometer scanner; FG, function generator; PC, personal computer.}
	\label{fig:example} % give each figure a logical label name
\end{figure*}

\section*{Results}
\subsection*{Experimental System}

Our experimental system is referenced to the metrological time base provided by two phase-stabilized optical frequency combs, which provide inherently low femtosecond-level timing jitter \cite{Cundiff:2003}. Figure~1A shows the experimental implementation. The key point of this method is that each detected probe pulse is assigned a unique, deterministic pump--probe delay calculated from comb-defined pulse indices, and the data recorded in laboratory-time order are subsequently reordered into an ascending sequence along the comb-defined delay axis. Because a fully stabilized frequency-comb pair can maintain mutual timing jitter below 50~fs over tens of minutes \cite{Okano:2022}, the pulse indices remain sufficiently well defined to assign each detected probe pulse a unique comb-defined pump--probe delay and to reorder the acquired data accordingly.

The experiment involves three pulse trains: the probe comb with a repetition frequency of $f_{\mathrm{rep1}}$, the seed comb with a repetition frequency of $f_{\mathrm{rep2}}$, and the amplified pump train delivered to the sample. The seed comb does not directly excite the sample; rather, it provides a phase-coherent seed pulse train from which the regenerative amplifier selects one pulse out of every $m$ seed-comb pulses. The amplified pump train therefore has a repetition frequency of
\begin{align}
f_{\mathrm{repRA}} = \frac{f_{\mathrm{rep2}}}{m},
\end{align}
Thus, the seed comb defines the fine timing grid for the pump pulses, whereas the regenerative amplifier converts this grid into a down-counted, high-energy pump train.

Figures~1B and 1C summarize the timing and reordering concept. Once an RA pump pulse and a probe-comb pulse temporally overlap, the relative delay between subsequent RA pump pulses and the corresponding probe-comb pulses advances by a small delay increment $\Delta t_{\mathrm{step}}$ from one pump cycle to the next. This row-to-row delay shift accumulates over many RA pump cycles. After one full acquisition cycle in laboratory time, $ T_{\mathrm{all}}$, the accumulated shift reaches one probe-comb period, $1/f_{\mathrm{rep1}}$, thereby connecting adjacent delay blocks in Fig.~1C.

The signal is acquired row-wise in laboratory time, with each row containing one sample from each of $m$ delay blocks. Consecutive rows shift all delay blocks by $\Delta t_{\mathrm{step}}$, so the laboratory-time sequence contains samples from different parts of the final delay scan in an interleaved order. The data are then reordered by probe-pulse order within the pump period and concatenated column-wise to yield a uniformly sampled pump--probe trace. As a result, a transient waveform over one RA period,
\begin{align}
T_{\mathrm{RA}} = \frac{1}{f_{\mathrm{repRA}}},
\end{align}
is reconstructed without mechanically scanning the pump--probe delay. This reconstruction differs from conventional ASOPS: the laboratory-time signal is not directly a time-stretched replica of the pump--probe trace, but an interleaved sequence of comb-indexed samples that must be deinterleaved according to the probe-pulse order within each pump period. Details are provided in Materials and Methods.

\begin{figure*} % Do NOT use \begin{figure*}
	\centering
	\includegraphics[width=1\textwidth]{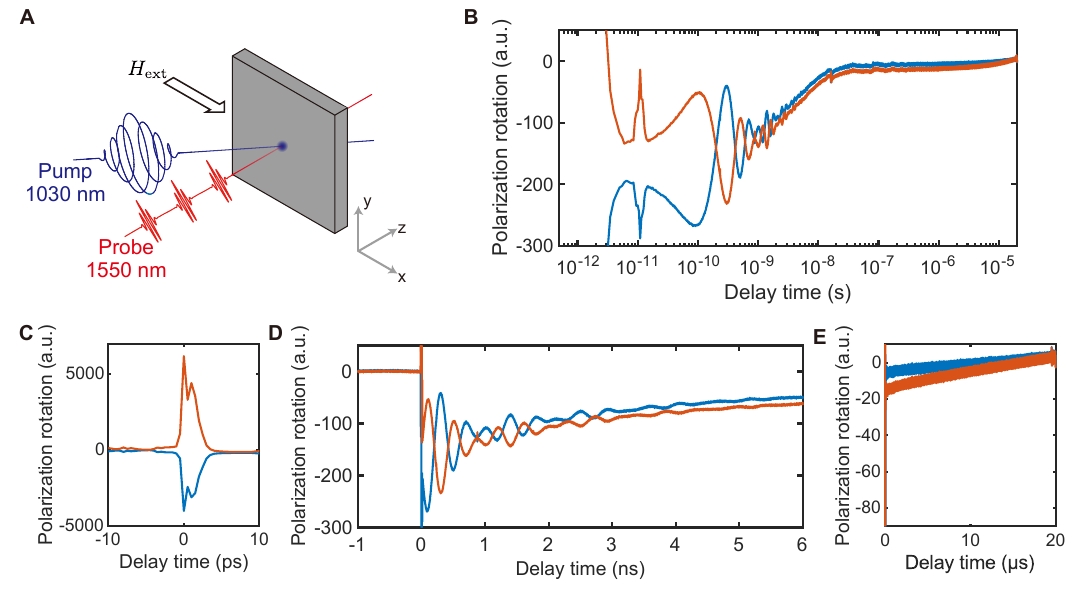} % for an image file named example_figure.*
	% Pick an appropriate width - in print, figures are usually one or two columns wide, which can
	% be approximated by 0.3\textwidth or 0.6\textwidth respectively. Use appropriate label sizes.

	% Captions go below figures
\caption{\textbf{Ultrawide-window pump--probe measurement of Bi-YIG.}
\textbf{(A)} Experimental setup.
\textbf{(B)} Probe polarization rotation measured at the pump--probe overlap position over the full 20~$\mu$s observation window on a logarithmic time axis. Blue and red traces correspond to $\sigma^+$ and $\sigma^-$ pump helicities, respectively.
\textbf{(C)} Enlarged view of the prompt response around $t=0$.
\textbf{(D)} Enlarged view of the nanosecond-scale magnetization precession.
\textbf{(E)} Same data as in (B) plotted on a linear time axis, showing the long-lived microsecond-scale relaxation.}
	\label{fig:example} % give each figure a logical label name
\end{figure*}

\subsection*{Multiscale magnon and phonon dynamics in Bi-YIG}

%Magnon and phonon dynamics provide a stringent test system for ultrawide-window pump--probe spectroscopy because they span impulsive femtosecond excitation, picosecond coherent motion, spatial propagation, boundary interactions, and microsecond relaxation. Ultrafast optical excitation can induce demagnetization and coherent spin motion \cite{Beaurepaire:1996, Kampen:2002, Kimel:2004, Kimel:2005, Hansteen:2006, Bigot:2009, Kirilyuk:2010}, launch coherent phonons \cite{Thomsen:1984, Zeiger:1992, Merlin:1997, Ruello:2015}, and subsequently drive relaxation through magnetic damping, spin--lattice equilibration, and heat diffusion \cite{Kittel:1953, Spencer:1960, Sanders:1977, Cahill:2004, Schmidt:2008, Bonse:2018}. Magnetoelastic coupling and magnon--phonon hybridization further provide pathways for energy exchange and mode conversion \cite{Kittel:1958, Scherbakov:2010, Kim:2012, Ogawa:2015, Hashimoto:2018, Hioki:2019, Hioki:2022}. However, previous measurements have generally accessed only limited portions of this broad temporal evolution \cite{Shimojima:2021, Hioki:2022, Hashimoto:2017, Nishikawa:2025}. Capturing these processes continuously requires a measurement platform that can connect the generation, propagation, repeated boundary reflection, and long-time relaxation of spin and lattice excitations on a single calibrated time axis.

To examine multiscale dynamics of magnons and phonons, we performed pump--probe measurements on a Bi-YIG sample under an in-plane magnetic field of 90~mT (Fig.~2A). Circularly polarized 1025-nm pump pulses were used to excite the sample, while linearly polarized 1550-nm probe pulses were transmitted through the sample to measure the pump-induced polarization rotation. The pump and probe beams were first spatially overlapped on the sample to measure the local response at the excitation position. Details of the sample properties, laser parameters, polarization analysis, and acquisition conditions are provided in Materials and Methods.

Figure~2B shows the temporal evolution of the probe polarization rotation measured at the pump--probe overlap position. The response is plotted on a logarithmic time axis from 500~fs to 20~$\mu$s and contains $40\,000\,575$ delay points assigned by the comb-defined time base. Immediately after pump excitation, an ultrafast polarization rotation appears at $t=0$. As shown in Fig.~2C, this prompt response reverses sign when the pump helicity is switched, consistent with an impulsive magneto-optical response induced by the inverse Faraday effect \cite{Kimel:2005, Satoh:2010}.

After the prompt response, an oscillatory polarization rotation persists for several nanoseconds, as shown in the magnified trace in Fig.~2D. The phase of this oscillation also reverses with pump helicity, indicating that the magnetization precession is launched by the helicity-dependent impulsive effective magnetic field generated by the circularly polarized pump pulse. Figure~2E shows the same response over the full 20~$\mu$s observation window on a linear time axis. In addition to the nanosecond-scale precession, the signal exhibits a slowly varying background that rises after photoexcitation and relaxes over the microsecond window. We attribute this long-lived component to pump-induced local heating and the subsequent release of thermally generated strain as heat dissipates from the excited region. The difference in the magnitude of the slope between the two circular polarization helicities is likely attributable to magnetic circular dichroism \cite{Scott:1975}. Specifically, the helicity-dependent absorption induced by magnetic circular dichroism leads to different temperature changes, resulting in different slopes. This microsecond-scale response motivates the spatiotemporal measurements below, in which heat-driven phonon propagation and repeated boundary reflections are directly imaged over the full observation window.

\begin{figure*} % Do NOT use \begin{figure*}
	\centering
	\includegraphics[width=1\textwidth]{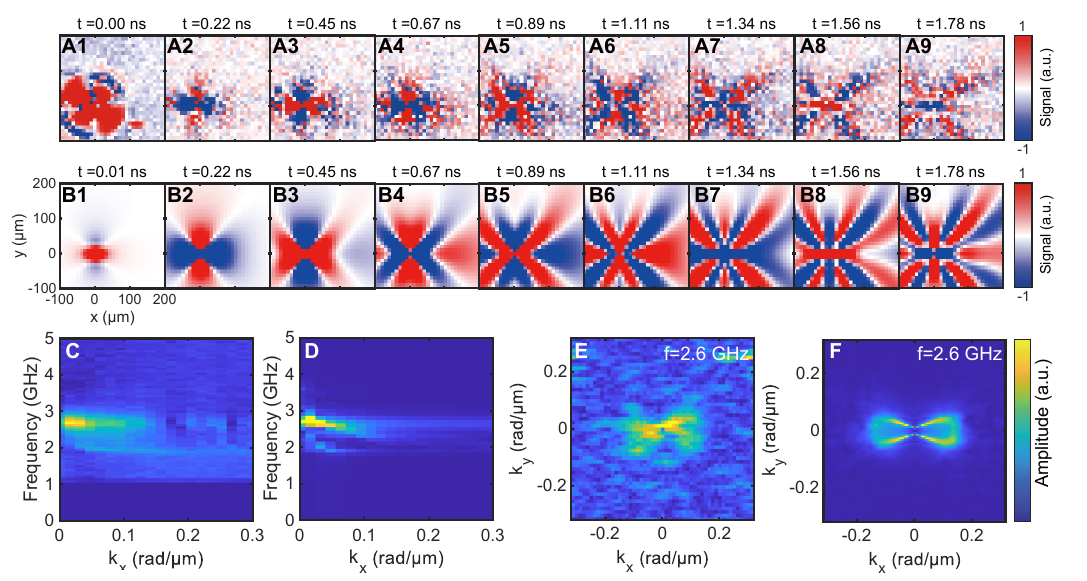} % for an image file named example_figure.*
	% Pick an appropriate width - in print, figures are usually one or two columns wide, which can
	% be approximated by 0.3\textwidth or 0.6\textwidth respectively. Use appropriate label sizes.

	% Captions go below figures
\caption{\textbf{Rapid high-density imaging of optically induced magnon propagation.}
\textbf{(A1--A9)} Spatiotemporal snapshots of the measured spin-wave propagation following optical excitation, obtained by two-dimensional scanning of the pump position. The incoherent thermal background has been subtracted.
\textbf{(B1--B9)} Corresponding micromagnetic simulation results under the same conditions.
\textbf{(C)} Spin-wave dispersion obtained from the measured data by a two-dimensional Fourier transform along the $x$ direction. The characteristic backward-volume magnetostatic spin-wave (BVMSW) behavior is reproduced.
\textbf{(D)} Simulated dispersion relation for comparison.
\textbf{(E)} Spin-wave amplitude at $f=2.6\,\mathrm{GHz}$ plotted in $(k_x,k_y)$ space, obtained from the full $(x,y,t)$ dataset.
\textbf{(F)} Corresponding reciprocal-space map from the simulation. The spectral intensity is concentrated along specific directions, indicating the anisotropic propagation of the spin waves.}
	\label{fig:example} % give each figure a logical label name
\end{figure*}

\subsection*{Rapid high-density imaging of magnon propagation}

A key advantage of the present system is that its rapid acquisition enables imaging measurements even when the dataset contains an exceptionally large number of temporal sampling points. To visualize the two-dimensional propagation of optically excited magnons, we performed the experiment while scanning the pump position over the sample surface. For the imaging measurements described below, we increased the detuning frequency to 137~Hz, corresponding to a temporal resolution of 44.6~ps and $448\,447$ sampling points, in order to accelerate the acquisition. Figures~3A1--A9 show representative snapshots of the measured spin-wave propagation after subtraction of the incoherent thermal background. The corresponding micromagnetic simulation results obtained under the same conditions are shown in Figs.~3B1--B9. A movie compiling the full time sequence up to 2~ns is provided as Supplementary Movie~1.

The measured spin-wave dynamics exhibit the characteristic behavior of backward-volume magnetostatic spin waves (BVMSWs). Along the $x$ axis, which is parallel to the applied magnetic field, the wavefronts propagate back toward the pump center. In addition, consistent with previous two-dimensional spin-wave imaging studies \cite{Satoh:2012}, the propagation is not isotropic but concentrated along specific directions. The experimentally observed spatiotemporal evolution agrees well with the micromagnetic simulations over the full delay range.

To further quantify the propagation, we extracted the time-resolved signal along the $x$ direction, i.e., parallel to the magnetic field, and obtained the spin-wave dispersion by applying a two-dimensional Fourier transform. The resulting dispersion relation (Fig.~3C) reproduces the characteristic BVMSW behavior, in which the frequency decreases with increasing wave number, and is in good agreement with the simulated dispersion shown in Fig.~3D.

We next Fourier-transformed the full $(x,y,t)$ dataset into $(k_x,k_y,f)$ space and plotted the spin-wave amplitude at $f=2.6\,\mathrm{GHz}$ on the $(k_x,k_y)$ plane (Fig.~3E). Here, $k_x$ and $k_y$ are the wave numbers along the $x$ and $y$ directions, respectively. The frequency of 2.6 GHz was selected as it corresponds to the center frequency of the propagating spin-wave packet. The corresponding map obtained from the simulations is shown in Fig.~3F. In both experiment and simulation, the spectral intensity is concentrated along specific directions characterized by $\phi=\tan^{-1}(k_y/k_x)$, making the directional nature of the spin-wave propagation evident also in reciprocal space \cite{Hashimoto:2017}.

\begin{figure*} % Do NOT use \begin{figure*}
	\centering
	\includegraphics[width=1\textwidth]{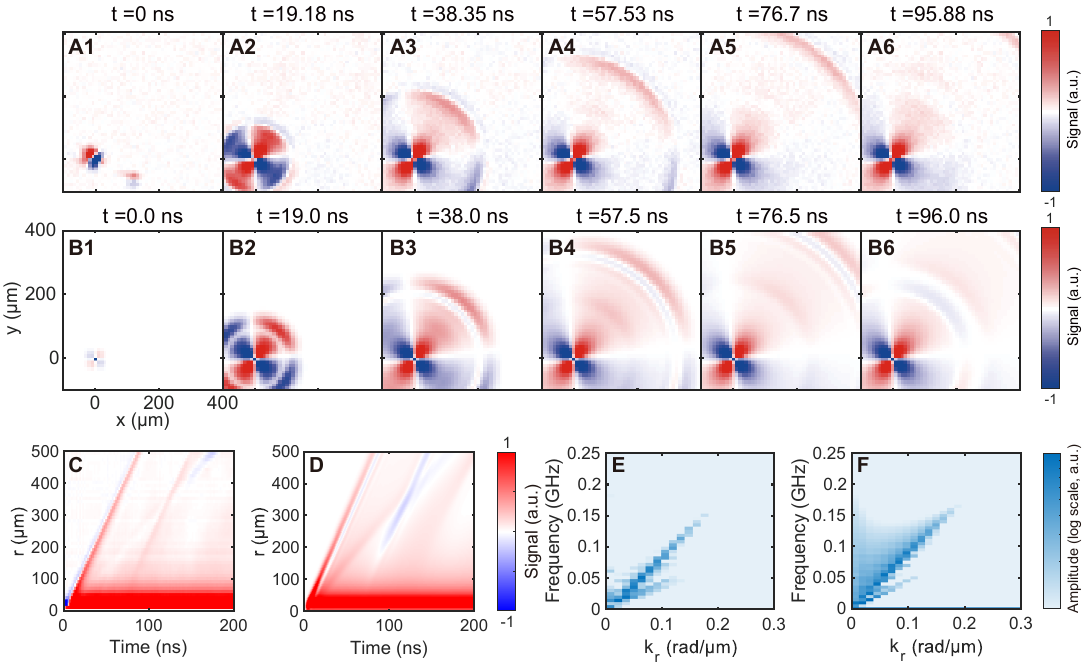} % for an image file named example_figure.*
	% Pick an appropriate width - in print, figures are usually one or two columns wide, which can
	% be approximated by 0.3\textwidth or 0.6\textwidth respectively. Use appropriate label sizes.

	% Captions go below figures
\caption{\textbf{Imaging of heat-driven acoustic phonon propagation in Bi-YIG.}
\textbf{(A1--A6)} Time-resolved images of the pump-induced polarization rotation at selected delay times from 0 to 95.88~ns. The acoustic wave packet expands outward from the pump-excited region with a quadrupolar angular contrast.
\textbf{(B1--B6)} Corresponding finite-element-method simulations of the photoelastic signal generated by laser-induced thermal expansion. The simulations reproduce both the outward propagation of the acoustic wavefront and the quadrupolar spatial pattern observed experimentally.
\textbf{(C)} Experimental spatiotemporal map of the acoustic-wave signal along the $r$ direction, showing propagation of the wavefront away from the excitation region.
\textbf{(D)} Simulated spatiotemporal map corresponding to (C).
\textbf{(E)} Experimental dispersion obtained by Fourier transforming the spatiotemporal data in (C).
\textbf{(F)} Simulated dispersion obtained from (D). The slopes of the upper and lower branches, assigned to the S0 and A0 modes, correspond to acoustic group velocities of 5.65 km/s and 2.48 km/s, respectively.}
	\label{fig:example} % give each figure a logical label name
\end{figure*}

\begin{figure*} % Do NOT use \begin{figure*}
	\centering
	\includegraphics[width=1\textwidth]{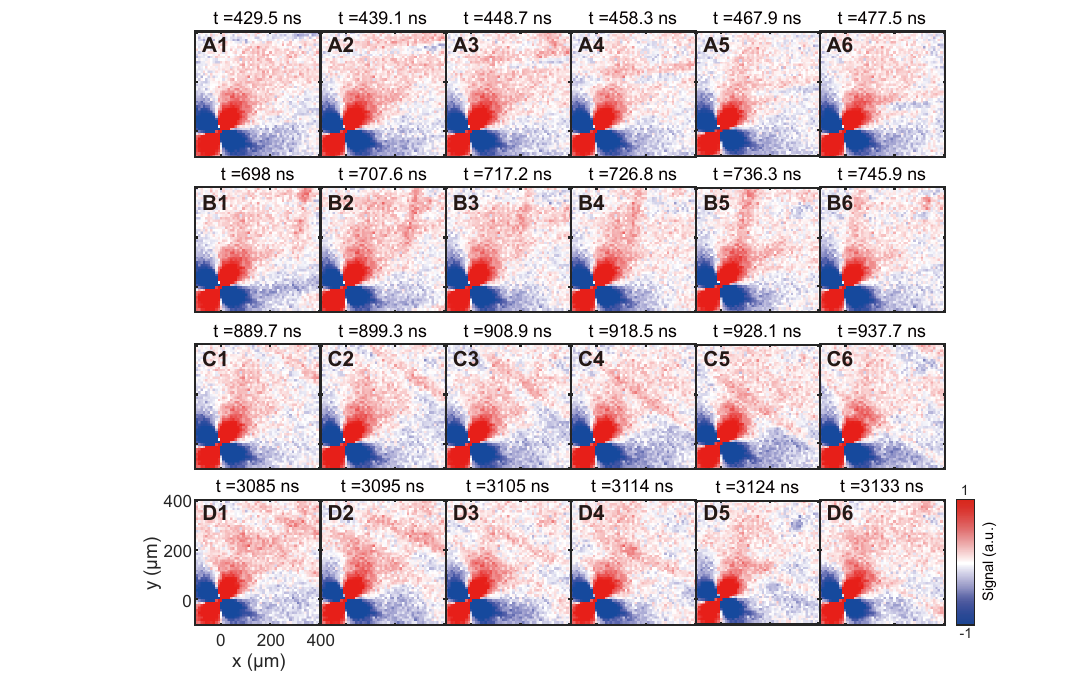} % for an image file named example_figure.*
	% Pick an appropriate width - in print, figures are usually one or two columns wide, which can
	% be approximated by 0.3\textwidth or 0.6\textwidth respectively. Use appropriate label sizes.

	% Captions go below figures
\caption{\textbf{Long-time imaging of edge-reflected acoustic waves in Bi-YIG.}
Selected time-resolved images of the pump-induced polarization rotation at representative long delay times. Reflected acoustic wavefronts returning from the sample edges are directly visualized. In panels A1--A6, the reflected wavefront propagates from top to bottom. In panels B1--B6, it propagates from right to left. In panels C1--C6, it moves diagonally from the upper right to the lower left, providing direct real-space evidence of successive reflections from different sample boundaries. Reflected waves remain clearly detectable even more than 3~$\mu$s after optical excitation, as seen in panels D1--D6. Repeated reflections generate increasingly complex strain patterns over the extended delay window; the full spatiotemporal evolution over 20~$\mu$s is shown in Supplementary Movie~2.}
	\label{fig:example} % give each figure a logical label name
\end{figure*}

\subsection*{Full-window imaging of heat-driven phonon propagation}

The magnon images in the previous section focused on the first 2~ns after photoexcitation, but the same acquisition provides the full pump--probe delay range up to 20~$\mu$s. This full-window capability is essential for imaging slower lattice dynamics: even at a delay increment of 44.6~ps, the raw trace contains approximately $4.5\times10^{5}$ delay points at each spatial position, yet the acquisition remains fast enough to perform two-dimensional imaging. At longer delays, we observed polarization-rotation signals arising from elastic waves launched by the optical pump pulse. A time-lapse movie of the polarization rotation over the entire 20~$\mu$s range is provided as Supplementary Movie~2. Because the elastic-wave signal evolves much more slowly than the spin-wave signal and becomes weaker at long delays, the raw data were binned into 1.9-ns intervals and averaged within each bin to improve the signal-to-noise ratio.

We first examine the early elastic-wave propagation up to 100~ns. Figures~4A1--A6 show snapshots of the measured polarization rotation. A wave packet expands outward from the excitation spot with a nearly circular wavefront and a pronounced quadrupolar angular dependence. %From the temporal evolution of the wavefront position, we estimate a group velocity of 5.7~km/s, comparable to the acoustic velocity of elastic waves in Bi-substituted YIG\cite{Siu:2001}.
The quadrupolar contrast originates from photoelastic modulation of the probe polarization by the strain field associated with the elastic wave. A dominant magnetoelastic contribution is unlikely because the signal does not reverse sign when the externally applied magnetic field is reversed. Within the photoelastic picture, the polarization-rotation signal detected by balanced detection is proportional to the imaginary part of the off-diagonal optical response \cite{Saito:2010},
\begin{align}
    \Delta I\propto \mathrm{Im}(Z_{xy})
\end{align}
\begin{align}
    Z_{xy}=\frac{ik^2}{2k_0}\int^d_0{\varepsilon_{\mathrm{pe},xy}}dz'
\end{align}
\begin{align}
    \varepsilon_{\mathrm{pe},xy}=P_{44}\eta_{xy}=2\left(\eta_{rr}-\frac{u_r}{r}\right)\sin{2\theta}.
\end{align}
Here, $k$ and $k_0$ are the optical wavenumbers in the sample and in vacuum, respectively; $\varepsilon_{\mathrm{pe},xy}$ is the photoelastic change in the off-diagonal component of the permittivity tensor; $P_{44}$ is the shear photoelastic coefficient; $\eta_{xy}$ is the shear strain; $\eta_{rr}$ is the radial strain; $u_r$ is the radial displacement; $r$ is the radial coordinate; and $\theta$ is the angle measured from the probe-polarization axis. The $\sin 2\theta$ factor naturally gives rise to a quadrupolar spatial pattern. The slight rotation of the experimental pattern is attributed to a small misalignment of the probe-polarization axis, estimated to be $6^\circ$.

To test this interpretation quantitatively, we performed finite-element-method (FEM) simulations of the strain field generated by laser-induced thermal expansion. The calculated strain distribution was converted into the expected photoelastic signal by integrating the off-diagonal photoelastic response along the sample thickness and by including the measured $6^\circ$ probe-polarization tilt. The resulting snapshots, shown in Figs.~4B1--B6, reproduce the experimental images well, including both the outward wavefront expansion and the quadrupolar contrast. The simulations also show that the quadrupolar contrast near the pump-irradiated region remains after the propagating wave packet has passed. This residual component indicates that the strain generation is not purely impulsive but contains a step-like contribution associated with local thermal expansion.
Supplementary Movie~3 provides the full time sequence up to 200 ns, showing a comparison of the experimental and calculated results.

To further analyze the observed elastic waves, the $(x,y,t)$ data were converted into radial $(r,t)$ maps by correcting the angular dependence of the signal with the $\sin 2\theta$ factor and subsequently integrating over the azimuthal angle, excluding angular regions where $|\sin 2\theta |$ is close to zero. The experimentally obtained radial map is shown in Fig. 4C, and the corresponding FEM simulation result is shown in Fig. 4D. In both maps, two wave packets with different propagation velocities are visible, suggesting the excitation of two distinct guided acoustic modes.
We assign these two modes to the fundamental symmetric and antisymmetric Lamb-wave modes, S0 and A0, respectively \cite{Nakamura:2021,Nakamura:2023}. The faster branch is attributed to the S0 mode, whereas the slower branch is attributed to the A0 mode. A second S0-like wave packet is also observed at a delay of approximately 60 ns. Since this delayed wave packet has nearly the same slope as the first S0-like wave packet in the $(r,t)$ map, it can be interpreted as the same Lamb-wave mode launched at a later time.
%The origin of this delayed S0-like wave packet is likely related to acoustic dynamics along the thickness direction. For the present sample thickness of 225 μm and a sound velocity of 5.6 km/s, the one-way and round-trip acoustic transit times are approximately 40 and 80 ns, respectively. Therefore, the observed delay of ∼60 ns does not simply correspond to either the one-way or round-trip propagation time of a pure longitudinal acoustic pulse. This suggests that the delayed S0-like emission may arise from a more complex coupling between thickness-direction acoustic motion, including reflected or standing-wave-like components, and the in-plane Lamb-wave mode. 
%We note that the phase of the second S0-like wave packet is reversed between the experiment and the simulation; the origin of this discrepancy is currently unclear.

Figures 4E and 4F show the corresponding dispersion curves obtained by applying a two-dimensional Fourier transform to the $(r,t)$ data in Figs. 4C and 4D, respectively. Note that the residual strain component was removed before applying the two-dimensional Fourier transform. In the $(k_r,f)$ space, two branches are clearly resolved. The upper branch is assigned to the S0 mode, while the lower branch is assigned to the A0 mode. From the slopes of these branches, the effective group velocities of the S0 and A0 modes were estimated to be 5.65 km/s and 2.48 km/s, respectively. The higher velocity of the S0 branch and its proximity to the longitudinal sound velocity of Bi-substituted YIG \cite{Siu:2001}, together with the slower and more dispersive character of the A0 branch, support the assignment of the observed modes to the fundamental symmetric and antisymmetric Lamb-wave modes.

At still longer delays, the ultrawide window provides direct access to long-lived spatiotemporal strain profiles associated with acoustic phonons. Most notably, reflected acoustic wavefronts returning from the sample edges are clearly visualized in Fig.~5. Their motion can be followed directly in real space, revealing successive reflections from different sample boundaries and the resulting redirection of the propagating strain wave. These reflected waves remain readily detectable even more than 3~$\mu$s after optical excitation, demonstrating that the present method can continuously track acoustic propagation, boundary reflection, and subsequent evolution over the microsecond regime. The strain modulations generated by repeated reflections persist throughout the entire 20~$\mu$s observation window, producing increasingly complex wave patterns, as shown in Supplementary Movie~2. 
%Because the pump spot was scanned in the present imaging scheme, the observed reflected-wave profiles should not be interpreted as simple fixed-source wavefront images. Nevertheless, the systematic motion of these features with delay time and their redirection at the sample boundaries indicate that they originate from edge-reflected acoustic waves.

Finally, at the longest delay times ($>1~\mu$s), the gradual weakening of the quadrupolar contrast near the pump-irradiated region is observed in Fig. 2E and Supplementary Movie~2, which is attributed to relaxation of the pump-induced temperature rise and the accompanying release of residual thermal strain. Because Bi-YIG is an electrically insulating dielectric, electronic heat transport is absent; lateral thermal diffusion is therefore expected to be much slower than the elastic-wave propagation over the present 20~$\mu$s timescale.

\section*{Discussion}

In this work, we have demonstrated ultralong pump–probe spectroscopy of magnon and phonon dynamics extending up to 20 $\mu$s with 500 fs temporal resolution. The resulting densely sampled movies, comprising $4.5 \times 10^5$ frames, enable us to visualize the evolution of spin and lattice excitations from their impulsive optical generation as coherent spin waves and elastic waves to their propagation, elastic-wave reflection at sample boundaries, and long-time relaxation in Bi-YIG. This capability for long-timescale measurements arises from the stability and deterministic nature of the optical-frequency-comb time base.

Such long-duration, densely sampled pump--probe imaging would be particularly advantageous for detecting long-lived magnons and phonons \cite{Hioki:2022, Serha:2026,Shao:2019} that persist beyond the time windows accessible with conventional pump--probe measurements. Nevertheless, our system currently provides a measurement time window of 20 $\mu$s, limiting its applicability to quasiparticles with longer lifetimes. Using low-repetition-rate pump laser, the technique could readily be extended to pump--probe movies over millisecond-scale and longer time windows while maintaining a comparable temporal grid. This extension would enable imaging of ultralong-lived phonons \cite{MacCabe:2020} as well as thermally excited magnons associated with the intrinsic spin Seebeck effect \cite{Jamison:2019}. Moreover, because this level of dense sampling over such long time windows has not previously been accessible, the method could serve not only as a diagnostic tool for known quasiparticle dynamics but also as a means of uncovering time--hidden metastable states that may escape detection in coarse temporal scans \cite{Li:2025}.

Overall, our method translates the metrological precision of optical frequency combs into a practical platform for ultrafast pump--probe movie over a microsecond-scale temporal range with picosecond-scale resolution. The approach should be applicable to a wide range of nonequilibrium phenomena in which electronic, spin, lattice, thermal, and structural dynamics evolve across widely separated timescales. By bridging the gap between ultrafast excitation and long-time relaxation, this technique provides a route to observing nonequilibrium dynamics as continuous processes, offering insights relevant to future information-processing devices based on long-lived quasiparticles.

\section*{Materials and Methods}

\subsection*{Experimental setup and delay-time reordering calibration}

The laser system was based on a dual-comb platform combined with a regenerative amplifier. The dual comb consisted of two home-built, fully stabilized, mode-locked erbium-fiber lasers, denoted as Comb~1 and Comb~2~\cite{Nishikawa:2023}. The two combs were phase-locked via a narrow-linewidth external-cavity continuous-wave laser (PLANEX, Redfern Integrated Optics), and the carrier-envelope offset frequencies of both combs were also phase-locked to a \(22~\mathrm{MHz}\) radio-frequency reference. The repetition rates of Comb~1 and Comb~2 are denoted by \(f_{\mathrm{rep1}}\) and \(f_{\mathrm{rep2}}\), respectively, and the detuning frequency is defined as
\begin{align}
\Delta f_{\mathrm{rep}}=f_{\mathrm{rep2}}-f_{\mathrm{rep1}}>0 .
\label{EqA1}
\end{align}
In the present system, \(f_{\mathrm{rep1}}\) and \(f_{\mathrm{rep2}}\) were approximately \(61.605~\mathrm{MHz}\).
$\Delta f_{\mathrm{rep}}$ was set around 1 -- 200 Hz depending on the condition as discussed later.

Comb~1 was amplified with an erbium-doped fiber amplifier and used as the \(1550~\mathrm{nm}\) probe comb. Comb~2 was amplified with an erbium-doped fiber amplifier, spectrally broadened to \(1025~\mathrm{nm}\) in a highly nonlinear fiber, and further amplified with a ytterbium-doped fiber amplifier to seed the regenerative amplifier (CPA-20, Kokyo). Thus, Comb~1 defined the probe-pulse train and the digitizer sampling clock, whereas Comb~2 defined the seed-pulse timing for the amplified pump train. The regenerative amplifier selected one pulse out of every \(m\) pulses of Comb~2. Therefore, the repetition rate of the amplified pump pulse train was
\begin{align}
f_{\mathrm{rep,RA}}=\frac{f_{\mathrm{rep2}}}{m}.
\label{EqA2}
\end{align}
In this experiment, \(m=1232\), giving \(f_{\mathrm{rep,RA}}\simeq 50.004~\mathrm{kHz}\). The corresponding pump-to-pump observation window in effective time was
\begin{align}
T_{\mathrm{RA}}=\frac{1}{f_{\mathrm{rep,RA}}}\simeq 20~\mu\mathrm{s}.
\label{EqA3}
\end{align}

The delay-time axis was reconstructed from the deterministic timing relationship among the probe comb, the seed comb, and the down-counted pump pulse train. The lock conditions were chosen such that
\begin{align}
M=\frac{f_{\mathrm{rep1}}}{\Delta f_{\mathrm{rep}}},
\qquad
M+1=\frac{f_{\mathrm{rep2}}}{\Delta f_{\mathrm{rep}}},
\label{EqA4}
\end{align}
were integers. Under this condition, after a laboratory time
\begin{align}
T_{\mathrm{all}}=\frac{1}{\Delta f_{\mathrm{rep}}},
\label{EqA5}
\end{align}
Comb~2 has advanced by exactly one pulse relative to Comb~1. Therefore, the same comb-to-comb timing pattern starts again.

The detuning frequency and lock conditions were chosen so that
\begin{align}
M=mp-1,
\label{EqA6}
\end{align}
with integer \(p\). Therefore, from Eqs. (\ref{EqA3}) -- (\ref{EqA6}), one obtains,
\begin{align}
T_{\mathrm{all}}
=
\frac{mp-1}{f_{\mathrm{rep1}}}
=
\frac{mp}{f_{\mathrm{rep2}}}
=
pT_{\mathrm{RA}} .
\label{EqA7}
\end{align}
Thus, one cycle corresponds to \(p\) RA pump periods, whereas the detector signal is sampled by Comb~1 and contains \(M=mp-1\) independent sampling points in a single nonoverlapping trace.

In laboratory time, the detector output is recorded as a one-dimensional sequence of Comb~1-clocked samples at the sampling rate \(f_{\mathrm{rep1}}\). For indexing this acquisition sequence, we introduce a nominal \(p\times m\) grid and write the laboratory-time sample index as
\begin{align}
q=k+jm,
\qquad
k=0,\ldots,m-1,\quad j=0,\ldots,p-1 .
\label{EqA8}
\end{align}
Here, \(k\) labels the order of the Comb~1-clocked probe sample within each block, and \(j\) labels successive sample blocks associated with successive RA pump pulses. Thus, in laboratory time, \(k\) changes first at fixed \(j\), and the data are acquired row-wise in the nominal grid.

This laboratory-time acquisition order is different from the effective pump--probe delay-time order. The delay shift between adjacent rows at fixed \(k\) is defined as
\begin{align}
\Delta t_{\mathrm{step}}
&=
\frac{m}{f_{\mathrm{rep1}}}
-
\frac{m}{f_{\mathrm{rep2}}}
\nonumber\\
&=
\frac{m\Delta f_{\mathrm{rep}}}
{f_{\mathrm{rep1}}f_{\mathrm{rep2}}}
=
\frac{\Delta f_{\mathrm{rep}}}
{f_{\mathrm{rep1}}f_{\mathrm{rep,RA}}}.
\label{EqA9}
\end{align}
This expression represents the difference between the laboratory-time interval of \(m\) Comb~1-clocked probe samples, \(m/f_{\mathrm{rep1}}\), and the interval between successive RA pump pulses, \(m/f_{\mathrm{rep2}}=1/f_{\mathrm{rep,RA}}\). Thus, although the RA pump pulses associated with neighboring rows are separated by one pump period \(T_{\mathrm{RA}}\) in laboratory time, the assigned pump--probe delay at fixed \(k\) shifts by only \(\Delta t_{\mathrm{step}}\) from one row to the next in the effective-time representation.

Taking the first pump--probe temporal overlap as \(\tau=0\), the delay assigned to the \((j,k)\)-th grid element in effective time is
\begin{align}
\tau_{j,k}
=
j\Delta t_{\mathrm{step}}
+
\frac{k}{f_{\mathrm{rep1}}}.
\label{EqA10}
\end{align}
Using Eqs. (\ref{EqA5}), (\ref{EqA7}), and (\ref{EqA9}), one obtains,
$
p\Delta t_{\mathrm{step}}
=
1/f_{\mathrm{rep1}}.
$
Therefore, Eq.~(\ref{EqA10}) can be rewritten as
\begin{align}
\tau_{j,k}
=
\left(j+kp\right)\Delta t_{\mathrm{step}}.
\label{EqA12}
\end{align}
The effective-time index is then defined as
\begin{align}
n=j+kp .
\label{EqA13}
\end{align}
The reordered trace is obtained by sorting the laboratory-time samples according to this effective-time index. This sorting corresponds to reading the nominal grid column-wise, in contrast to the row-wise acquisition in laboratory time.

Because one nonoverlapping timing cycle contains $M=mp-1$ independent Comb~1-clocked samples, only
\begin{align}
q=0,\ldots,mp-2
\end{align}
are included in a single trace. The terminal grid point \(q=mp-1\), corresponding to \((j,k)=(p-1,m-1)\), is not counted as an independent data point because it is equivalent to the first point of the next timing cycle. In effective time, this terminal point would correspond to
\begin{align}
\tau_{p-1,m-1}
=
(mp-1)\Delta t_{\mathrm{step}}
=
T_{\mathrm{RA}} .
\end{align}
Thus, one effective pump--probe trace consists of
\begin{align}
n=0,\ldots,mp-2
\end{align}
delay points and spans the half-open observation window
\begin{align}
0\le \tau<T_{\mathrm{RA}} .
\end{align}

The detuning frequency \(\Delta f_{\mathrm{rep}}\) determines the trade-off between delay increment and acquisition time. At fixed \(f_{\mathrm{rep1}}\) and \(f_{\mathrm{rep,RA}}\), decreasing \(\Delta f_{\mathrm{rep}}\) increases $p$
and decreases $\Delta t_{\mathrm{step}}$.
Therefore, a smaller \(\Delta f_{\mathrm{rep}}\) gives a finer delay increment and denser sampling along the effective time axis. At the same time, the laboratory-frame acquisition period $T_{\mathrm{all}}$ increases.
The detuning frequency was therefore chosen by balancing the required delay increment against the acquisition time for one complete reordered trace.

For the long-window measurements, \(\Delta f_{\mathrm{rep}}\) was set to \(1.54~\mathrm{Hz}\). With \(f_{\mathrm{rep1}}=61.605~\mathrm{MHz}\) and \(f_{\mathrm{rep,RA}}=50.004~\mathrm{kHz}\), this gives
\[
\Delta t_{\mathrm{step}}
=
\frac{\Delta f_{\mathrm{rep}}}{f_{\mathrm{rep1}}f_{\mathrm{rep,RA}}}
\simeq
500~\mathrm{fs}.
\]
The effective observation window was \(T_{\mathrm{RA}}=20~\mu\mathrm{s}\), yielding approximately \(40\,000\,575\) delay points in one reordered trace. For imaging measurements requiring faster acquisition, \(\Delta f_{\mathrm{rep}}\) was increased to approximately \(137~\mathrm{Hz}\), giving a delay increment of \(44.6~\mathrm{ps}\) and approximately \(448\,447\) delay points over the same \(20~\mu\mathrm{s}\) observation window.

\subsection*{Measurements of magnon and phonon dynamics}

The sample was a commercial Bi-substituted iron-garnet Faraday rotator designed for operation at \(1.55~\mu\mathrm{m}\) (GLB, Granopt). It was a liquid-phase-epitaxy-grown single crystal with a thickness of \(225~\mu\mathrm{m}\) and lateral dimensions of \(5\times5~\mathrm{mm}^2\). According to the manufacturer's specifications, the Faraday rotation angle at \(1550~\mathrm{nm}\) was \(45^{\circ}\), and the saturation magnetization was \(4\pi M_s=1250~\mathrm{G}\). Ferromagnetic resonance measurements were performed using an ESR spectrometer (ELEXSYS E500, Bruker) at \(f_{\mathrm{FMR}}=9.86~\mathrm{GHz}\) under an external magnetic field of \(H_{\mathrm{ext}}=3100~\mathrm{Oe}\). From the full width at half maximum of the resonance line, \(\Delta H=70~\mathrm{Oe}\), the Gilbert damping constant was estimated as \(\alpha_G\approx \Delta H/(2H_{\mathrm{ext}})=0.01\).

In the pump--probe measurements, the output of the regenerative amplifier (\(1025~\mathrm{nm}\), \(400~\mathrm{fs}\), \(0.6~\mu\mathrm{J}\) per pulse) was converted to circular polarization with a quarter-wave plate and used as the pump beam. The probe beam was taken from Comb~1 at \(1550~\mathrm{nm}\), with a pulse energy of \(0.2~\mathrm{nJ}\), and was linearly polarized. The pump and probe beams were focused onto the sample with spot diameters of \(50~\mu\mathrm{m}\) and \(30~\mu\mathrm{m}\), respectively. These conditions corresponded to fluences of \(31~\mathrm{mJ\,cm^{-2}}\) for the pump and \(28~\mu\mathrm{J\,cm^{-2}}\) for the probe. An external magnetic field of \(90~\mathrm{mT}\) was applied along the in-plane \(x\) direction to saturate the magnetization.

For imaging measurements, the probe position was fixed, whereas the pump spot was scanned over a \(500\times500~\mu\mathrm{m}^2\) area in \(10~\mu\mathrm{m}\) steps using a galvanometric mirror. After transmission through the sample, the probe polarization was analyzed using a quarter-wave plate and a polarizing beam splitter. The two polarization components were then detected with a balanced photoreceiver (HBPR-450M-10K-IN-FST, FEMTO).

The detector output was digitized at the sampling rate \(f_{\mathrm{rep1}}\) using a digitizer (M2p.5962-x4, Spectrum). The sampling clock was obtained from the synchronizer output of the signal generator (WF1968, NF Corp.) used in the \(f_{\mathrm{rep1}}\)-locking loop. The phase of the signal generator was adjusted so that the Comb~1-clocked sampling timing coincided with the peak of the detector response. The raw data were recorded in laboratory-time order and converted to the effective pump--probe delay trace using the pulse-index reordering procedure described above. 

For the multiscale magnon and phonon dynamics measurement shown in Fig.~2, approximately 1000 acquisitions were averaged. The reordered waveform contained non-sample-related background components and reordering-induced artifacts. A continuous-wave-like oscillatory background appeared at the probe repetition frequency \(f_{\mathrm{rep1}}\) and its harmonics; these spectral components were removed by applying notch filters after Fourier transformation. In addition, short pulse-like artifacts recurring every \(16~\mathrm{ns}\), approximately corresponding to \(1/f_{\mathrm{rep1}}\), were observed in the time-domain waveform. Except for the prompt response at \(\tau=0\), these pulse-like features were treated as outliers and replaced by interpolation between neighboring valid data points.

For the propagation imaging measurements shown in Figs.~3--5, approximately 1000 acquisitions were averaged at each spatial position. For visualization of the slow acoustic-wave dynamics in Figs.~4 and 5, the reordered traces were further binned into \(1.9~\mathrm{ns}\) intervals and averaged within each bin to improve the signal-to-noise ratio.

\subsection*{Finite-element simulations}

Finite-element simulations were performed using COMSOL Multiphysics 6.4. Spin-wave simulations were carried out with the Micromagnetics Module \cite{Lan:2015} coupled to the AC/DC Module. The magnetization dynamics were described by the Landau--Lifshitz--Gilbert equation,
\[
\frac{d\mathbf{M}}{dt}
=
-\gamma\,\mathbf{M}\times\mathbf{H}_{\mathrm{eff}}
+
\frac{\alpha_G}{M_s}\,\mathbf{M}\times\frac{d\mathbf{M}}{dt},
\]
where \(\mathbf{M}\) is the magnetization, \(\gamma=28~\mathrm{GHz/T}\) is the gyromagnetic ratio, and \(\mathbf{H}_{\mathrm{eff}}=\mathbf{H}_{\mathrm{ext}}+\mathbf{H}_{\mathrm{demag}}+\mathbf{H}_{\mathrm{IFE}}\) is the effective magnetic field including the external field, the demagnetizing field, and the inverse-Faraday-effect field. The optical excitation was modeled as
\[
\mathbf{H}_{\mathrm{IFE}}
=
H_{\mathrm{IFE}}
\exp\!\left(-\frac{t^2}{\tau^2}\right)
\exp\!\left(-\frac{2(x^2+y^2)}{\sigma^2}\right),
\]
where \(H_{\mathrm{IFE}}\) is the peak field amplitude, and \(\sigma\) is the pump-spot radius. The pump pulse width $\tau$ only needs to be sufficiently short compared to the spin-wave period. Therefore, for computational convenience, we set $\tau=5 \,\mathrm{ps}$. The demagnetizing field was obtained by solving
\[
\nabla^2 V_m=\nabla\cdot\mathbf{M},
\qquad
\mathbf{H}_{\mathrm{demag}}=-\nabla V_m,
\]
where \(V_m\) is the magnetic scalar potential.

Elastic-wave simulations were performed with the Heat Transfer Module and the Solid Mechanics Module in COMSOL Multiphysics, following the thermoelastic finite-element approach used in previous simulations of photoinduced strain dynamics in thin plates~\cite{Nakamura:2021}. In the present study, the model was adapted to a two-dimensional axisymmetric $(r,z)$ geometry, because optically excited elastic waves propagate approximately isotropically from the pump-irradiated region. The simulation domain was a rectangle of width \(2.5~\mathrm{mm}\) and height \(225~\mu\mathrm{m}\), revolved around the \(z\) axis. The laser-induced temperature rise was calculated from
\[
C\frac{\partial T}{\partial t}-K_T\nabla^2 T=P(r,z,t),
\]
where \(T\) is the temperature, \(C\) is the heat capacity, \(K_T\) is the thermal conductivity, and \(P(r,z,t)\) is the heat source term,
%\begin{widetext}
\[
P(r,z,t)
=
P_0
\exp\!\left(-\frac{t^2}{\tau^{\,2}}\right)
\exp\!\left(-\frac{2r^2}{\sigma^{\,2}}\right)
\exp\!\left(\frac{z-d}{\ell_{\mathrm{opt}}}\right).
\]
%\end{widetext}
Here, $P_0$ is the heat-source amplitude, \(\ell_{\mathrm{opt}}=139~\mu\mathrm{m}\) is the optical penetration depth, and \(d\) is the sample thickness. The resulting thermoelastic stress,
\[
\sigma_{\mathrm{TE}}=-3B\beta\left(T-T_0\right),
\]
was introduced into the Navier--Cauchy equation,
\[
\rho\frac{\partial^2\mathbf{u}}{\partial t^2}
=
\mu\nabla^2\mathbf{u}
+
(\lambda+\mu)\nabla(\nabla\cdot\mathbf{u})
+
\nabla\sigma_{\mathrm{TE}},
\]
where \(\mathbf{u}\) is the displacement, \(\rho\) is the mass density, and \(\lambda\) and \(\mu\) are the Lam\'e parameters. To compare the simulation with the experimentally observed shear strain, \(\eta_{xy}=2(\eta_{rr}-u_r/r)\sin 2\theta\), we exported \((\eta_{rr}-u_r/r)\) from COMSOL and multiplied it by \(\sin 2\theta\) in MATLAB.

\section*{Acknowledgments}
We acknowledge Prof. Y. Ota and Dr. S. Gao for providing the Bi-YIG sample used in this study.
Parts of this study are supported by JSPS KAKENHI (23K26165, 24K21743, 24H01202, 25KJ2061, 25H02153, 26H02144), JST CREST (Grant No. JPMJCR19J4), MEXT Q-LEAP (Grant No. JPMXS0118067246), and Precise Measurement Technology Promotion Foundation. 
ChatGPT (OpenAI, GPT-5.5) was used for language refinement and improvement of clarity and readability. All scientific content, interpretation, and conclusions were developed and verified by the authors.

\section*{Author contributions}
S. W. conceived the experiments and supervised the project. R. S. performed experimental measurement and finite-element simulation of the elastic wave. S. F. performed the finite-element simulation of the spin wave.  R. S. and S. W. wrote the manuscript and all the authors contributed to constructive discussions.
% 	\section*{Author contributions}
% S. F. designed and supervised the project. S.F and K.W. performed the experimental measurement. S.F. conducted the finite-element simulation. K. W. fabricated the crystalline microresonators, and S. F., K. W., and S. K. developed the experimental setups.  S. F wrote the manuscript with input from T. T. and all the authors contributed to constructive discussions.
%\clearpage
% 	\bibliographystyle{nature}			
	\bibliography{science_template}% Produces the bibliography via BibTeX.

@article{Satoh:2010,
   author = {Satoh, Takuya and Cho, Sung-Jin and Iida, Ryugo and Shimura, Tsutomu and Kuroda, Kazuo and Ueda, Hiroaki and Ueda, Yutaka and Ivanov, B. A. and Nori, Franco and Fiebig, Manfred},
   title = {Spin Oscillations in Antiferromagnetic {NiO} Triggered by Circularly Polarized Light},
   journal = {Phys. Rev. Lett.},
   volume = {105},
   number = {7},
   pages = {077402},
   DOI= {10.1103/PhysRevLett.105.077402},
   year = {2010},
   type = {Journal Article}
}

@article{Li:2025,
   author = {Li, Xinwei and Esin, Iliya and Han, Youngjoon and Liu, Yincheng and Zhao, Hengdi and Ning, Honglie and Barrett, Cora and Shan, Jun-Yi and Seyler, Kyle and Cao, Gang and Refael, Gil and Hsieh, David},
   title = {Time-hidden magnetic order in a multi-orbital Mott insulator},
   journal = {Nat. Phys.},
   volume = {21},
   number = {3},
   pages = {451-457},
   ISSN = {1745-2481},
   DOI = {10.1038/s41567-024-02752-1},
   url = {https://doi.org/10.1038/s41567-024-02752-1
https://www.nature.com/articles/s41567-024-02752-1.pdf},
   year = {2025},
   type = {Journal Article}
}

@article{Satoh:2012,
   author = {Satoh, T. and Terui, Y. and Moriya, R. and Ivanov, B. A. and Ando, K. and Saitoh, E. and Shimura, T. and Kuroda, K.},
   title = {Directional control of spin-wave emission by spatially shaped light},
   journal = {Nat. Photon.},
   volume = {6},
   number = {10},
   pages = {662-666},
   DOI= {10.1038/nphoton.2012.218},
   year = {2012},
   type = {Journal Article}
}

@article{Saito:2010,
   author = {Saito, T. and Matsuda, O. and Tomoda, M. and Wright, O. B.},
   title = {Imaging gigahertz surface acoustic waves through the photoelastic effect},
   journal = {J. Opt. Soc. Am. B},
   volume = {27},
   number = {12},
   pages = {2632-2638},
   DOI= {10.1364/josab.27.002632},
   year = {2010},
   type = {Journal Article}
}

@article{Kimel:2005,
   author = {Kimel, A. V. and Kirilyuk, A. and Usachev, P. A. and Pisarev, R. V. and Balbashov, A. M. and Rasing, {Th.}},
   title = {Ultrafast non-thermal control of magnetization by instantaneous photomagnetic pulses},
   journal = {Nature},
   volume = {435},
   number = {7042},
   pages = {655-657},
   DOI = {10.1038/nature03564},
   url = {https://doi.org/10.1038/nature03564},
   year = {2005},
   type = {Journal Article}
}

@article{Ogawa:2015,
   author = {Ogawa, Naoki and Koshibae, Wataru and Beekman, Aron Jonathan and Nagaosa, Naoto and Kubota, Masashi and Kawasaki, Masashi and Tokura, Yoshinori},
   title = {Photodrive of magnetic bubbles via magnetoelastic waves},
   journal = {Proc. Natl. Acad. Sci. U.S.A.},
   volume = {112},
   number = {29},
   pages = {8977-8981},
   DOI = {10.1073/pnas.1504064112},
   url = {https://doi.org/10.1073/pnas.1504064112
https://www.pnas.org/doi/pdf/10.1073/pnas.1504064112},
   year = {2015},
   type = {Journal Article}
}

@article{Hashimoto:2018,
   author = {Hashimoto, Y. and Bossini, D. and Johansen, T. H. and Saitoh, E. and Kirilyuk, A. and Rasing, T.},
   title = {Frequency and wavenumber selective excitation of spin waves through coherent energy transfer from elastic waves},
   journal = {Phys. Rev. B},
   volume = {97},
   number = {14},
   pages = {140404},   
   year = {2018},
   type = {Journal Article},
   ISSN = {2469-9950},
   DOI = {10.1103/PhysRevB.97.140404},
   url = {https://journals.aps.org/prb/pdf/10.1103/PhysRevB.97.140404}
}

@article{Hioki:2019,
   author = {Hioki, Tomosato and Hashimoto, Yusuke and Johansen, Tom H. and Saitoh, Eiji},
   title = {Time-Resolved Imaging of Magnetoelastic Waves by the Cotton-Mouton Effect},
   journal = {Phys. Rev. Appl.},
   volume = {11},
   number = {6},
   pages = {061007},
   DOI = {10.1103/PhysRevApplied.11.061007},
   url = {https://doi.org/10.1103/PhysRevApplied.11.061007
https://journals.aps.org/prapplied/pdf/10.1103/PhysRevApplied.11.061007},
   year = {2019},
   type = {Journal Article}
}

@article{Hioki:2022,
   author = {Hioki, T. and Hashimoto, Y. and Saitoh, E.},
   title = {Coherent oscillation between phonons and magnons},
   journal = {Commun. Phys.},
   volume = {5},
   number = {1},
   pages = {115},
   ISSN = {2399-3650},
   DOI = {10.1038/s42005-022-00888-1},
   url = {<Go to ISI>://WOS:000793841500001
https://www.nature.com/articles/s42005-022-00888-1.pdf},
   year = {2022},
   type = {Journal Article}
}

@article{Hashimoto:2017,
   author = {Hashimoto, Yusuke and Daimon, Shunsuke and Iguchi, Ryo and Oikawa, Yasuyuki and Shen, Ka and Sato, Koji and Bossini, Davide and Tabuchi, Yutaka and Satoh, Takuya and Hillebrands, Burkard and Bauer, Gerrit E. W. and Johansen, Tom H. and Kirilyuk, Andrei and Rasing, Theo and Saitoh, Eiji},
   title = {All-optical observation and reconstruction of spin wave dispersion},
   journal = {Nat. Commun.},
   volume = {8},
   pages = {15859},
   DOI = {10.1038/ncomms15859},
   url = {https://doi.org/10.1038/ncomms15859
https://www.nature.com/articles/ncomms15859.pdf},
   year = {2017},
   type = {Journal Article}
}

@article{Nishikawa:2025,
   author = {Nishikawa, Daichi and Maezawa, Kazuki and Shibata, Riku and Fujii, Shun and Watanabe, Shinichi},
   title = {Ultrafast movies of the photoinduced phonon and magnon propagation using dual frequency-comb technology},
   journal = {Opt. Lett.},
   volume = {50},
   number = {6},
   pages = {1929-1932},
   DOI = {10.1364/OL.553529},
   url = {https://opg.optica.org/ol/abstract.cfm?URI=ol-50-6-1929},
   year = {2025},
   type = {Journal Article}
}

@article{Yao:2025,
   author = {Yao, Yihang and Hao, Danyang and Zhang, Qicheng},
   title = {Perspectives on Devices for Integrated Phononic Circuits},
   journal = {Adv. Mater.},
   volume = {37},
   number = {23},
   pages = {2407642},
   ISSN = {0935-9648},
   DOI = {https://doi.org/10.1002/adma.202407642},
   url = {https://doi.org/10.1002/adma.202407642},
   year = {2025},
   type = {Journal Article}
}

@article{Elzinga:1987a,
   author = {Elzinga, P. A. and Lytle, F. E. and Jian, Y. and King, G. B. and Laurendeau, N. M.},
   title = {Pump/probe spectroscopy by asynchronous optical-sampling},
   journal = {Appl. Spectrosc.},
   volume = {41},
   number = {1},
   pages = {2-4},
   DOI= {10.1366/0003702874868025},
   year = {1987},
   type = {Journal Article}
}

@article{Elzinga:1987b,
   author = {Elzinga, P. A. and Kneisler, R. J. and Lytle, F. E. and Jiang, Y. and King, G. B. and Laurendeau, N. M.},
   title = {Pump/probe method for fast analysis of visible spectral signatures utilizing asynchronous optical sampling},
   journal = {Appl. Opt.},
   volume = {26},
   number = {19},
   pages = {4303-4309},
   DOI= {10.1364/ao.26.004303},
   year = {1987},
   type = {Journal Article}
}

@article{Yasui:2005,
   author = {Yasui, T. and Saneyoshi, E. and Araki, T.},
   title = {Asynchronous optical sampling terahertz time-domain spectroscopy for ultrahigh spectral resolution and rapid data acquisition},
   journal = {Appl. Phys. Lett.},
   volume = {87},
   number = {6},
   pages = {061101},
   DOI= {10.1063/1.2008379},
   year = {2005},
   type = {Journal Article}
}

@article{Bartels:2006,
   author = {Bartels, A. and Hudert, F. and Janke, C. and Dekorsy, T. and K{\"o}hler, K.},
   title = {Femtosecond time-resolved optical pump-probe spectroscopy at kilohertz-scan-rates over nanosecond-time-delays without mechanical delay line},
   journal = {Appl. Phys. Lett.},
   volume = {88},
   number = {4},
   pages = {041117},
   DOI= {10.1063/1.2167812},
   year = {2006},
   type = {Journal Article}
}

@article{Krauss:2015,
   author = {Krau{\ss}, N. and Sch{\"a}fer, G. and Flock, J. and Kliebisch, O. and Li, C. and Barros, H. G. and Heinecke, D. C. and Dekorsy, T.},
   title = {Two-colour high-speed asynchronous optical sampling based on offset-stabilized {Yb:KYW} and {Ti:sapphire} oscillators},
   journal = {Opt. Express},
   volume = {23},
   number = {14},
   pages = {18288-18299},
   DOI= {10.1364/oe.23.018288},
   year = {2015},
   type = {Journal Article}
}

@article{Asahara:2020,
   author = {Asahara, A. and Arai, Y. and Saito, T. and Ishi-Hayase, J. and Akahane, K. and Minoshima, K.},
   title = {Dual-comb-based asynchronous pump-probe measurement with an ultrawide temporal dynamic range for characterization of photo-excited InAs quantum dots},
   journal = {Appl. Phys. Express},
   volume = {13},
   number = {6},
   pages = {062003},
   DOI= {10.35848/1882-0786/ab8b4f},
   year = {2020},
   type = {Journal Article}
}

@article{Okano:2022,
   author = {Okano, Makoto and Watanabe, Shinichi},
   title = {Triggerless data acquisition in asynchronous optical-sampling terahertz time-domain spectroscopy based on a dual-comb system},
   journal = {Opt. Express},
   volume = {30},
   number = {22},
   pages = {39613-39623},
   DOI= {10.1364/OE.472192},
   year = {2022},
   type = {Journal Article}
}

@article{Velsink:2023,
   author = {Velsink, M. C. and Illienko, M. and Sudera, P. and Witte, S.},
   title = {Optimizing pump-probe reflectivity measurements of ultrafast photoacoustics with modulated asynchronous optical sampling},
   journal = {Rev. Sci. Instrum.},
   volume = {94},
   number = {10},
   pages = {103002},
   DOI= {10.1063/5.0155006},
   year = {2023},
   type = {Journal Article}
}

@article{Flory:2023,
   author = {Flöry, Tobias and Stummer, Vinzenz and Pupeikis, Justinas and Willenberg, Benjamin and Nussbaum-Lapping, Alexander and Kaksis, Edgar and Camargo, Franco V. A. and Barkauskas, Martynas and Phillips, Christopher R. and Keller, Ursula and Cerullo, Giulio and Pugžlys, Audrius and Baltuška, Andrius},
   title = {Rapid-Scan Nonlinear Time-Resolved Spectroscopy over Arbitrary Delay Intervals},
   journal = {Ultrafast Sci.},
   volume = {3},
   pages = {0027},
   DOI = {10.34133/ultrafastscience.0027},
   year = {2023},
   type = {Journal Article}
}

@article{Domke:2018,
   author = {Domke, Matthias and Wick, Sebastian and Laible, Maike and Rapp, Stephan and Huber, Heinz P. and Sroka, Ronald},
   title = {Ultrafast dynamics of hard tissue ablation using femtosecond-lasers},
   journal = {Journal of Biophotonics},
   volume = {11},
   number = {10},
   pages = {e201700373},
   ISSN = {1864-063X},
   DOI = {https://doi.org/10.1002/jbio.201700373},
   url = {https://doi.org/10.1002/jbio.201700373},
   year = {2018},
   type = {Journal Article}
}

@article{Hatanaka:2023,
   author = {Hatanaka, D. and Asano, M. and Okamoto, H. and Yamaguchi, H.},
   title = {Phononic Crystal Cavity Magnomechanics},
   journal = {Phys. Rev. Appl.},
   volume = {19},
   number = {5},
   pages = {054071},
   DOI = {10.1103/PhysRevApplied.19.054071},
   year = {2023},
   type = {Journal Article}
}

@article{He:2026,
   author = {He, Wenqing and Wan, Caihua and Lin, Jianing and Wu, Hao and Yu, Guoqiang and Han, Xiufeng},
   title = {Ultrafast spin-orbit-torque switching in a ferrimagnetic insulator with high compensation temperature},
   journal = {Phys. Rev. Appl.},
   volume = {25},
   number = {3},
   pages = {034092},
   DOI = {10.1103/6wcw-vmrj},
   year = {2026},
   type = {Journal Article}
}

@article{MacCabe:2020,
   author = {MacCabe, Gregory S. and Ren, Hengjiang and Luo, Jie and Cohen, Justin D. and Zhou, Hengyun and Sipahigil, Alp and Mirhosseini, Mohammad and Painter, Oskar},
   title = {Nano-acoustic resonator with ultralong phonon lifetime},
   journal = {Science},
   volume = {370},
   number = {6518},
   pages = {840-843},
   DOI = {10.1126/science.abc7312},
   url = {https://doi.org/10.1126/science.abc7312},
   year = {2020},
   type = {Journal Article}
}

@article{Jamison:2019,
   author = {Jamison, John S. and Yang, Zihao and Giles, Brandon L. and Brangham, Jack T. and Wu, Guanzhong and Hammel, P. Chris and Yang, Fengyuan and Myers, Roberto C.},
   title = {Long lifetime of thermally excited magnons in bulk yttrium iron garnet},
   journal = {Phys. Rev. B},
   volume = {100},
   number = {13},
   pages = {134402},
   DOI = {10.1103/PhysRevB.100.134402},
   year = {2019},
   type = {Journal Article}
}

@article{Kim:2018,
   author = {Kim, JunWoo and Cho, Byungmoon and Yoon, Tai Hyun and Cho, Minhaeng},
   title = {Dual-Frequency Comb Transient Absorption: Broad Dynamic Range Measurement of Femtosecond to Nanosecond Relaxation Processes},
   journal = {J. Phys. Chem. Lett.},
   volume = {9},
   number = {8},
   pages = {1866-1871},
   DOI= {10.1021/acs.jpclett.8b00886},
   year = {2018},
   type = {Journal Article}
}

@article{Bredenbeck:2004,
   author = {Bredenbeck, Jens and Helbing, Jan and Hamm, Peter},
   title = {Continuous scanning from picoseconds to microseconds in time resolved linear and nonlinear spectroscopy},
   journal = {Rev. Sci. Instrum.},
   volume = {75},
   number = {11},
   pages = {4462-4466},
   DOI= {10.1063/1.1793891},
   year = {2004},
   type = {Journal Article}
}

@article{Yu:2005,
   author = {Yu, Anchi and Ye, Xiong and Ionascu, Dan and Cao, Wenxiang and Champion, Paul M.},
   title = {Two-color pump-probe laser spectroscopy instrument with picosecond time-resolved electronic delay and extended scan range},
   journal = {Rev. Sci. Instrum.},
   volume = {76},
   number = {11},
   pages = {114301},
   DOI= {10.1063/1.2126808},
   year = {2005},
   type = {Journal Article}
}

@article{Nishikawa:2023,
   author = {Nishikawa, D. and Maezawa, K. and Fujii, S. and Okano, M. and Watanabe, S.},
   title = {A two-color dual-comb system for time-resolved measurements of ultrafast magnetization dynamics using triggerless asynchronous optical sampling},
   journal = {Rev. Sci. Instrum.},
   volume = {94},
   number = {6},
   pages = {063003},
   DOI= {10.1063/5.0147899},
   year = {2023},
   type = {Journal Article}
}

@article{Carroll:2009,
   author = {Carroll, E. C. and Hill, M. P. and Madsen, D. and Malley, K. R. and Larsen, D. S.},
   title = {A single source femtosecond-millisecond broadband spectrometer},
   journal = {Rev. Sci. Instrum.},
   volume = {80},
   number = {2},
   pages = {026102},
   DOI= {10.1063/1.3070516},
   year = {2009},
   type = {Journal Article}
}

@article{Nakagawa:2016,
   author = {Nakagawa, Tatsuo and Okamoto, Kido and Hanada, Hiroaki and Katoh, Ryuzi},
   title = {Probing with randomly interleaved pulse train bridges the gap between ultrafast pump-probe and nanosecond flash photolysis},
   journal = {Opt. Lett.},
   volume = {41},
   number = {7},
   pages = {1498-1501},
   DOI= {10.1364/OL.41.001498},
   year = {2016},
   type = {Journal Article}
}

@article{Helbing:2023,
   author = {Helbing, Jan and Hamm, Peter},
   title = {Versatile Femtosecond Laser Synchronization for Multiple-Timescale Transient Infrared Spectroscopy},
   journal = {J. Phys. Chem. A},
   volume = {127},
   number = {30},
   pages = {6347-6356},
   DOI= {10.1021/acs.jpca.3c03526},
   year = {2023},
   type = {Journal Article}
}

@article{Antonucci:2015,
   author = {Antonucci, Laura and Bonvalet, Adeline and Solinas, Xavier and Daniault, Louis and Joffre, Manuel},
   title = {Arbitrary-detuning asynchronous optical sampling with amplified laser systems},
   journal = {Opt. Express},
   volume = {23},
   number = {21},
   pages = {27931-27940},
   DOI= {10.1364/OE.23.027931},
   year = {2015},
   type = {Journal Article}
}

@article{Solinas:2017,
   author = {Solinas, Xavier and Antonucci, Laura and Bonvalet, Adeline and Joffre, Manuel},
   title = {Multiscale control and rapid scanning of time delays ranging from picosecond to millisecond},
   journal = {Opt. Express},
   volume = {25},
   number = {15},
   pages = {17811-17819},
   DOI= {10.1364/OE.25.017811},
   year = {2017},
   type = {Journal Article}
}

@article{Abbas:2014,
   author = {Abbas, A. and Guillet, Y. and Rampnoux, J. M. and Rigail, P. and Mottay, E. and Audoin, B. and Dilhaire, S.},
   title = {Picosecond time resolved opto-acoustic imaging with 48 {MHz} frequency resolution},
   journal = {Opt. Express},
   volume = {22},
   number = {7},
   pages = {7831-7843},
   DOI= {10.1364/oe.22.007831},
   year = {2014},
   type = {Journal Article}
}

@article{Cundiff:2003,
   author = {Cundiff, Steven T. and Ye, Jun},
   title = {Colloquium: Femtosecond optical frequency combs},
   journal = {Rev. Mod. Phys.},
   volume = {75},
   pages = {325-342},
   DOI= {10.1103/RevModPhys.75.325},
   year = {2003},
   type = {Journal Article}
}

@article{Siu:2001,
  title = {Magnons and acoustic phonons in {${\mathrm{Y}}_{3\ensuremath{-}x}{\mathrm{Bi}}_{x}{\mathrm{Fe}}_{5}{\mathrm{O}}_{12}$}},
  author = {Siu, G. G. and Lee, C. M. and Liu, Yulong},
  journal = {Phys. Rev. B},
  volume = {64},
  issue = {9},
  pages = {094421},
  numpages = {10},
  year = {2001},
  month = {Aug},
  publisher = {American Physical Society},
  doi = {10.1103/PhysRevB.64.094421},
  url = {https://link.aps.org/doi/10.1103/PhysRevB.64.094421}
}

@article{Scott:1975,
  title = {Magnetic circular dichroism and Faraday rotation spectra of {${\mathrm{Y}}_{3}$${\mathrm{Fe}}_{5}$${\mathrm{O}}_{12}$}},
  author = {Scott, G. B. and Lacklison, D. E. and Ralph, H. I. and Page, J. L.},
  journal = {Phys. Rev. B},
  volume = {12},
  issue = {7},
  pages = {2562--2571},
  numpages = {0},
  year = {1975},
  month = {Oct},
  publisher = {American Physical Society},
  doi = {10.1103/PhysRevB.12.2562},
  url = {https://link.aps.org/doi/10.1103/PhysRevB.12.2562}
}

@article{Lan:2015,
  title = {Spin-Wave Diode},
  author = {Lan, Jin and Yu, Weichao and Wu, Ruqian and Xiao, Jiang},
  journal = {Phys. Rev. X},
  volume = {5},
  issue = {4},
  pages = {041049},
  numpages = {7},
  year = {2015},
  month = {Dec},
  publisher = {American Physical Society},
  doi = {10.1103/PhysRevX.5.041049},
  url = {https://link.aps.org/doi/10.1103/PhysRevX.5.041049}
}

@article{Nakamura:2021,
   author = {A Nakamura and T Shimojima and K Ishizaka},
   doi = {10.1063/4.0000059},
   issn = {2329-7778},
   issue = {2},
   journal = {Structural Dynamics},
   month = {4},
   pages = {024103},
   title = {Finite-element simulation of photoinduced strain dynamics in silicon thin plates},
   volume = {8},
   url = {https://doi.org/10.1063/4.0000059},
   year = {2021}
}

@article{Nakamura:2023,
   author = {Asuka Nakamura and Takahiro Shimojima and Kyoko Ishizaka},
   doi = {10.1021/acs.nanolett.2c03938},
   issn = {1530-6984},
   issue = {7},
   journal = {Nano Lett.},
   month = {4},
   pages = {2490-2495},
   publisher = {American Chemical Society},
   title = {Characterizing an Optically Induced Sub-micrometer Gigahertz Acoustic Wave in a Silicon Thin Plate},
   volume = {23},
   url = {https://doi.org/10.1021/acs.nanolett.2c03938},
   year = {2023}
}

@article{Au:2013,
  title = {Direct Excitation of Propagating Spin Waves by Focused Ultrashort Optical Pulses},
  author = {Au, Y. and Dvornik, M. and Davison, T. and Ahmad, E. and Keatley, P. S. and Vansteenkiste, A. and Van Waeyenberge, B. and Kruglyak, V. V.},
  journal = {Phys. Rev. Lett.},
  volume = {110},
  issue = {9},
  pages = {097201},
  numpages = {5},
  year = {2013},
  month = {Feb},
  publisher = {American Physical Society},
  doi = {10.1103/PhysRevLett.110.097201},
  url = {https://link.aps.org/doi/10.1103/PhysRevLett.110.097201}
}

@article{Sugawara:2002,
  title = {Watching Ripples on Crystals},
  author = {Sugawara, Y. and Wright, O. B. and Matsuda, O. and Takigahira, M. and Tanaka, Y. and Tamura, S. and Gusev, V. E.},
  journal = {Phys. Rev. Lett.},
  volume = {88},
  issue = {18},
  pages = {185504},
  numpages = {4},
  year = {2002},
  month = {Apr},
  publisher = {American Physical Society},
  doi = {10.1103/PhysRevLett.88.185504},
  url = {https://link.aps.org/doi/10.1103/PhysRevLett.88.185504}
}

@article{Casals:2020,
  title = {Generation and Imaging of Magnetoacoustic Waves over Millimeter Distances},
  author = {Casals, Blai and Statuto, Nahuel and Foerster, Michael and Hern\'andez-M\'{\i}nguez, Alberto and Cichelero, Rafael and Manshausen, Peter and Mandziak, Ania and Aballe, Luc\'{\i}a and Hern\`andez, Joan Manel and Maci\`a, Ferran},
  journal = {Phys. Rev. Lett.},
  volume = {124},
  issue = {13},
  pages = {137202},
  numpages = {6},
  year = {2020},
  month = {Apr},
  publisher = {American Physical Society},
  doi = {10.1103/PhysRevLett.124.137202},
  url = {https://link.aps.org/doi/10.1103/PhysRevLett.124.137202}
}

@article{Maezawa:2024,
  title = {Spatiotemporal visualization of a surface acoustic wave coupled to magnons across a submillimeter-long sample by pulsed laser interferometry},
  author = {Maezawa, Kazuki and Fujii, Shun and Yamanoi, Kazuto and Nozaki, Yukio and Watanabe, Shinichi},
  journal = {Phys. Rev. Appl.},
  volume = {21},
  issue = {4},
  pages = {044047},
  numpages = {13},
  year = {2024},
  month = {Apr},
  publisher = {American Physical Society},
  doi = {10.1103/PhysRevApplied.21.044047},
  url = {https://link.aps.org/doi/10.1103/PhysRevApplied.21.044047}
}

@article{Wang:2022,
   author = {Wang, Ji-Qian and Zhang, Zi-Dong and Yu, Si-Yuan and Ge, Hao and Liu, Kang-Fu and Wu, Tao and Sun, Xiao-Chen and Liu, Le and Chen, Hua-Yang and He, Cheng and Lu, Ming-Hui and Chen, Yan-Feng},
   title = {Extended topological valley-locked surface acoustic waves},
   journal = {Nat. Commun.},
   volume = {13},
   number = {1},
   pages = {1324},
   ISSN = {2041-1723},
   DOI = {10.1038/s41467-022-29019-8},
   url = {https://doi.org/10.1038/s41467-022-29019-8
https://www.ncbi.nlm.nih.gov/pmc/articles/PMC8921310/pdf/41467_2022_Article_29019.pdf},
   year = {2022},
   type = {Journal Article}
}

@article{Shao:2019,
  title = {Phononic Band Structure Engineering for High-Q Gigahertz Surface Acoustic Wave Resonators on Lithium Niobate},
  author = {Shao, Linbo and Maity, Smarak and Zheng, Lu and Wu, Lue and Shams-Ansari, Amirhassan and Sohn, Young-Ik and Puma, Eric and Gadalla, M.N. and Zhang, Mian and Wang, Cheng and Hu, Evelyn and Lai, Keji and Lon\ifmmode \check{c}\else \v{c}\fi{}ar, Marko},
  journal = {Phys. Rev. Appl.},
  volume = {12},
  issue = {1},
  pages = {014022},
  numpages = {8},
  year = {2019},
  month = {Jul},
  publisher = {American Physical Society},
  doi = {10.1103/PhysRevApplied.12.014022},
  url = {https://link.aps.org/doi/10.1103/PhysRevApplied.12.014022}
}

@article{Serha:2026,
author = {Rostyslav O. Serha  and Kaitlin H. McAllister  and Fabian Majcen  and Sebastian Knauer  and Timmy Reimann  and Carsten Dubs  and Gennadii A. Melkov  and Alexander A. Serga  and Vasyl S. Tyberkevych  and Andrii V. Chumak  and Dmytro A. Bozhko },
title = {Ultralong-living magnons in the quantum limit},
journal = {Sci. Adv.},
volume = {12},
number = {18},
pages = {eaee2344},
year = {2026},
doi = {10.1126/sciadv.aee2344},
}

@article{Domke:2012,
author = {Matthias Domke and Stephan Rapp and Michael Schmidt and Heinz P. Huber},
journal = {Opt. Express},
number = {9},
pages = {10330--10338},
publisher = {Optica Publishing Group},
title = {Ultrafast pump-probe microscopy with high temporal dynamic range},
volume = {20},
month = {Apr},
year = {2012},
url = {https://opg.optica.org/oe/abstract.cfm?URI=oe-20-9-10330},
doi = {10.1364/OE.20.010330},
}

@article{Kajiwara:2010,
   author = {Kajiwara, Y. and Harii, K. and Takahashi, S. and Ohe, J. and Uchida, K. and Mizuguchi, M. and Umezawa, H. and Kawai, H. and Ando, K. and Takanashi, K. and Maekawa, S. and Saitoh, E.},
   title = {Transmission of electrical signals by spin-wave interconversion in a magnetic insulator},
   journal = {Nature},
   volume = {464},
   number = {7286},
   pages = {262-266},
   ISSN = {1476-4687},
   DOI = {10.1038/nature08876},
   url = {https://doi.org/10.1038/nature08876
https://www.nature.com/articles/nature08876.pdf},
   year = {2010},
   type = {Journal Article}
}

@article{Zhang:2021,
   author = {Zhang, Jianyu and Chen, Mingfeng and Chen, Jilei and Yamamoto, Kei and Wang, Hanchen and Hamdi, Mohammad and Sun, Yuanwei and Wagner, Kai and He, Wenqing and Zhang, Yu and Ma, Ji and Gao, Peng and Han, Xiufeng and Yu, Dapeng and Maletinsky, Patrick and Ansermet, Jean-Philippe and Maekawa, Sadamichi and Grundler, Dirk and Nan, Ce-Wen and Yu, Haiming},
   title = {Long decay length of magnon-polarons in $\mathrm{BiFeO_3/La_{0.67}Sr_{0.33}MnO_3}$ heterostructures},
   journal = {Nat. Commun.},
   volume = {12},
   number = {1},
   pages = {7258},
   ISSN = {2041-1723},
   DOI = {10.1038/s41467-021-27405-2},
   url = {https://doi.org/10.1038/s41467-021-27405-2},
   year = {2021},
   type = {Journal Article}
}
\end{document}